\documentclass[twoside,11pt]{article}

\usepackage{cjp}
\usepackage{graphics}

\begin{document}


\title{Investigating the Physical Origin of Unconventional Low-Energy\\ 
Excitations and Pseudogap Phenomena in Cuprate Superconductors}

\author{N.-C. Yeh$^1$, C.-T. Chen$^1$, A. D. Beyer$^1$, and S.-I. Lee$^2$}

\address{$^1$Department of Physics, California Institute of Technology, Pasadena, CA 91125, USA\\
$^2$Pohang University of Science and Technology, Pohang 790-784, Republic of Korea}

\maketitle              

\begin{abstract}
We investigate the physical origin of unconventional low-energy
excitations in cuprate superconductors by considering the effect of coexisting 
competing orders (CO) and superconductivity (SC) and of quantum fluctuations 
and other bosonic modes on the low-energy charge excitation spectra. 
By incorporating both SC and CO in the bare Green's function and quantum 
phase fluctuations in the self-energy, we can consistently account for 
various empirical findings in both the hole- and electron-type 
cuprates, including the excess subgap quasiparticle density of states,
``dichotomy'' in the fluctuation-renormalized quasiparticle spectral density 
in momentum space, and the occurrence and magnitude of a low-energy pseudogap 
being dependent on the relative gap strength of CO and SC. Comparing these
calculated results with experiments of ours and others, we suggest
that there are two energy scales associated with the pseudogap phenomena,
with the high-energy pseudogap probably of magnetic origin and the low-energy
pseudogap associated with competing orders. 
\end{abstract}


\section{Introduction}

High-temperature superconducting cuprates are doped Mott insulators with strongly 
correlated electronic ground states manifested in forms of different competing orders 
(CO) besides superconductivity (SC)~\cite{Zhang97,Chakravarty01,Sachdev03,Kivelson03,LeePA06}. 
Phenomenologically, 
the presence of coexisting SC and CO in the ground state has a number of
important implications: 1) the occurrence of quantum criticality 
becomes a natural consequence of competing phases in the ground state~\cite{Yeh02a};
2) significant quantum fluctuations are expected as the result
of proximity to quantum criticality~\cite{Yeh05a,Yeh05b,Zapf05}; 
3) the low-energy excitations of coexisting SC and CO differ from conventional 
Bogoliubov quasiparticles of pure SC because of redistribution of the spectral 
weight between SC and CO~\cite{ChenCT06,ChenCT03}; 4) the presence of competing 
orders and strong quantum fluctuations gives rise to weakened superconducting 
stiffness and extreme type-II nature in the cuprates~\cite{Yeh05a,Yeh05b}; 
5) the relevant competing orders can vary in different cuprates, giving rise to 
non-universal physical properties~\cite{Yeh02a,Yeh01,ChenCT02}; 
6) the existence of competing orders and accompanied unconventional low-energy 
excitations are likely relevant to the occurrence of pseudogap phenomena~\cite{ChenCT06}. 
We further note that empirically the pseudogap phenomena generally refer to two 
different energy scales; the lower energy scale is correlated with 
the Nernst effect~\cite{LeePA06,WangY06,WangY05} and the suppressed
quasiparticle density of states observed above $T_c$, and is only found in 
hole-type cuprates; whereas the higher energy scale is present in both 
electron- and hole-type cuprates according to optical~\cite{Onose01,Gallais05} 
and neutron scattering~\cite{Yamada98,Motoyama06} experiments. 
As discussed in our recent work~\cite{ChenCT06} and further 
elaborated below in this paper, the lower-energy pseudogap may be associated 
with the competing order, whereas the higher-energy pseudogap seems to be 
of magnetic origin.

In this work, we consider the effect of competing orders and quantum phase
fluctuations on the quasiparticle spectral function and density of states 
(DOS) by incorporating both SC and CO in the bare Green's function and quantum 
phase fluctuations in the proper self-energy. Using realistic bandstructures
and physical parameters in our calculations and comparing the resulting 
spectra with experimental data on both hole- and electron-type cuprates, we 
find favorable comparison of the concept of coexisting CO and SC with a wide 
range of experimental phenomena~\cite{ChenCT06}, including the occurrence of excess subgap 
quasiparticle density of states (DOS)~\cite{ChenCT02}, spatial modulations 
in the low-temperature quasiparticle spectra that are unaccounted for by 
Bogoliubov quasiparticles alone~\cite{ChenCT03,Hoffman02,Vershinin04,McElroy05}, 
``dichotomy'' in the momentum-dependent quasiparticle spectral
function from ARPES~\cite{ZhouXJ04}, and the presence (absence) of the low-energy 
pseudogap~\cite{ChenCT02,Renner98,Kleefisch01} in the hole 
(electron)-type cuprates above the SC transition. We also 
conjecture that the high-energy pseudogap observed in optical~\cite{Onose01,Gallais05} 
and neutron scattering~\cite{Yamada98,Motoyama06} experiments on the cuprates
may be associated with dynamic spin fluctuations, which can be stabilized by 
an external field. In this context, we present experimental evidence for current-induced
pseudogap in electron-type cuprates~\cite{Yeh05a,Yeh05b}. Finally, we propose 
a unified phase diagram for all cuprates based on the findings of competing orders 
and two pseudogap energy scales. 

\section{Unconventional low-energy excitations in cuprate superconductors}

In order to consider the effect of coexisting CO and SC on the low-energy
excitations consistently, we begin by incorporating both CO and SC in the mean-field
Hamiltonian, and then include the quantum phase fluctuations associated
with the CO and SC phases in the proper self-energy to solve the Dyson's equation
self-consistently for the full Green's function at $T=0$ so as to obtain the
quasiparticle spectral density function $A(\textbf{k},\omega)$ and the density of 
states ${\cal N} (\omega)$, as detailed in Ref.~\cite{ChenCT06}. 
For electron-type cuprates, we consider the case of 
coexisting $s$-wave SC and charge-density waves (CDW)~\cite{Kivelson03}, 
whereas for hole-type cuprates, we assume coexisting $d$-wave SC and disorder-pinned 
spin-density waves (SDW)~\cite{Demler01,Polkovnikov02}. We also examine the possibility 
of $d$-density waves (DDW)~\cite{Chakravarty01} being the relevant CO in $d$-wave SC, 
and in all cases, we employ the realistic bandstructures and Fermi levels for 
given cuprates under consideration. Specifically, we consider the bare Green's 
function $G_0 (\textbf{k},\omega)$ associated with the mean-field Hamiltonian
\begin{equation}
{\cal H}_{MF} = {\cal H}_{\rm SC} + {\cal H}_{\rm CO} 
\label{eq:MFsSCCDW}
\end{equation}
where ${\cal H}_{\rm SC}$ is the BCS-like superconducting Hamiltonian for a given 
pairing potential $\Delta _{\rm SC} (\textbf{k})$, with $\Delta _{\rm SC} (\textbf{k}) 
= \Delta _{\rm SC}$ being independent of $\textbf{k}$ in the case of $s$-wave pairing and 
$\Delta _{\rm SC} (\textbf{k}) \approx \Delta _{\rm SC} \lbrack \cos k_x a - \cos k_y a 
\rbrack /2$ for $d_{x^2-y^2}$-wave pairing. Here $a$ is the lattice constant of the
two-dimensional CuO$_2$ layer, ${\cal H}_{\rm CO}$ is the mean-field CO 
Hamiltonian with the CO energy scale given by $V_{\rm CO}$ and the wave-vector 
of the CO (among CDW, SDW and DDW) given by $\textbf{Q}$~\cite{ChenCT06}. 
In the zero-temperature zero-field limit, the quantum phase fluctuations are 
dominated by the longitudinal phase fluctuations, which can be approximated by 
the one-loop velocity-velocity correlation in the proper self-energy 
$\Sigma ^{\ast}$~\cite{ChenCT06}. Thus, the full Green's function 
$G (\textbf{k},\tilde \omega)$ is determined self-consistently through the Dyson's 
equation:
\begin{equation}
G^{-1} (\textbf{k}, \tilde \omega) = G_0 ^{-1} (\textbf{k}, \omega) 
- \Sigma ^{\ast} (\textbf{q}, \tilde \omega), 
\label{eq:Dyson}
\end{equation}
where $\tilde \omega$ denotes the energy renormalized by the phase fluctuations. Equation 
(2) is solved self-consistently~\cite{ChenCT06,Beyer06} by first choosing an energy 
$\omega$, going over the $\textbf{k}$-values in the Brillouin zone by summing over
a finite phase space in $\textbf{q}$ near each $\textbf{k}$, and then finding the corresponding 
$\tilde \xi _{\textbf{k}}$, $\tilde \omega$ and $\tilde \Delta$ until the solution to the full
Green's function $G(\textbf{k}, \tilde \omega)$ converges using iteration method~\cite{Beyer06}. 
The converged Green's function yields the spectral density function $A( \textbf{k}, \omega ) \equiv 
- \rm{Im} \left[ G(\textbf{k}, \tilde \omega (\textbf{k},\omega)) \right] / \pi$ and the DOS 
${\cal N} (\omega) \equiv \sum _{\textbf{k}} A( \textbf{k}, \omega ) $. We also
confirm the conservation of spectral weight throughout the self-consistent calculations.

Using the aforementioned approach, we find that many important features in the quasiparticle 
DOS of both the hole- and electron-type cuprates of varying doping levels can be well
accounted for by a set of parameters $( \Delta _{\rm SC}, V _{\rm CO}, \eta )$, provided that the CO
specified couples well to the Bogoliubov quasiparticles of the cuprates near the Fermi level,
and $\eta$ denotes the magnitude of the quantum phase fluctuations specified in 
Ref.~\cite{ChenCT06}. As exemplified in Fig.~1, we compare the quasiparticle tunneling
spectra of three different cuprates of varying doping levels with theoretically calculated
quasiparticle DOS for (a) $\rm Bi_2Sr_2CaCu_2O_x$ (Bi-2212)~\cite{McElroy05}, (b) 
$\rm YBa_2Cu_3O_x$ (Y-123)~\cite{Yeh01,Wei98}, and (c) optimally doped 
$\rm Sr_{0.9}La_{0.1}CuO_2$ (La-112)~\cite{ChenCT02}. The corresponding SC gap
($\Delta _{\rm SC}$) and CO energy ($V_{\rm CO}$) as a function of the doping level 
are determined from the fitting and shown in Fig.~1(d). In contrast to the common
empirical practice that denotes the peak-to-peak or hump-to-hump energy differences  
as twice of the nominal SC gap values without considering CO, we find 
that our theoretical fitting to the quasiparticle DOS not only captures the primary features
of the tunneling spectra but also yields doping dependent $\Delta _{\rm SC}$ that better follows
the doping dependence of $T_c$, whereas the doping dependence of $V_{\rm CO}$ increases
with decreasing doping level. 

Another important finding of our calculations is to provide a natural explanation for
the presence and the absence of low-energy pseudogap in hole-type and electron-type 
cuprate superconductors, respectively. That is, generally if $V_{\rm CO} > \Delta _{\rm SC}$,
the quasiparticle spectra exhibit two sets of peaks at energies $\omega \sim 
\pm \Delta _{\rm eff}$ and $\omega \approx \pm \Delta _{\rm SC}$ where
$\Delta _{\rm eff} \equiv \sqrt{\Delta _{\rm SC} ^2 + V_{\rm CO} ^2}$ . The peaks at 
$\pm \Delta _{\rm SC}$ vanish at $T_c$ while those associated with $\pm \Delta _{\rm eff}$ 
remain finite at $\omega \approx \pm V_{\rm CO}$ above $T_c$ and become 
much broadened due to thermal fluctuations. Thus, these broadened features at 
$\omega \approx \pm V_{\rm CO}$ above $T_c$ can be referred to as the lower-energy pseudogap. 
In contrast, for $V_{\rm CO} < \Delta _{\rm SC}$ and under finite quantum
fluctuations, the quasiparticle spectra only exhibit one set of peaks at energies 
$\omega \approx \pm \Delta _{\rm eff} \sim \pm \Delta _{\rm SC}$ as exemplified in Fig.~1(c), 
and no discernible low-energy pseudogap can be observed above $T_c$. We therefore
conclude that in under- and optimally doped hole-type cuprates, the condition 
$V_{\rm CO} > \Delta _{\rm SC}$ generally holds, whereas in electron-type cuprates, we
generally find $V_{\rm CO} < \Delta _{\rm SC}$. The disparity in the strength of the competing
energy scales may be related to the much better electronic coupling with the longitudinal 
optical phonons~\cite{Tachiki03} for the hole-type electronic configuration along the 
Cu-O bonding direction in the CuO$_2$ plane, as discussed in Refs.~\cite{Yeh05a,ChenCT06}. 
 
\begin{figure}[!t]
\centerline{\includegraphics{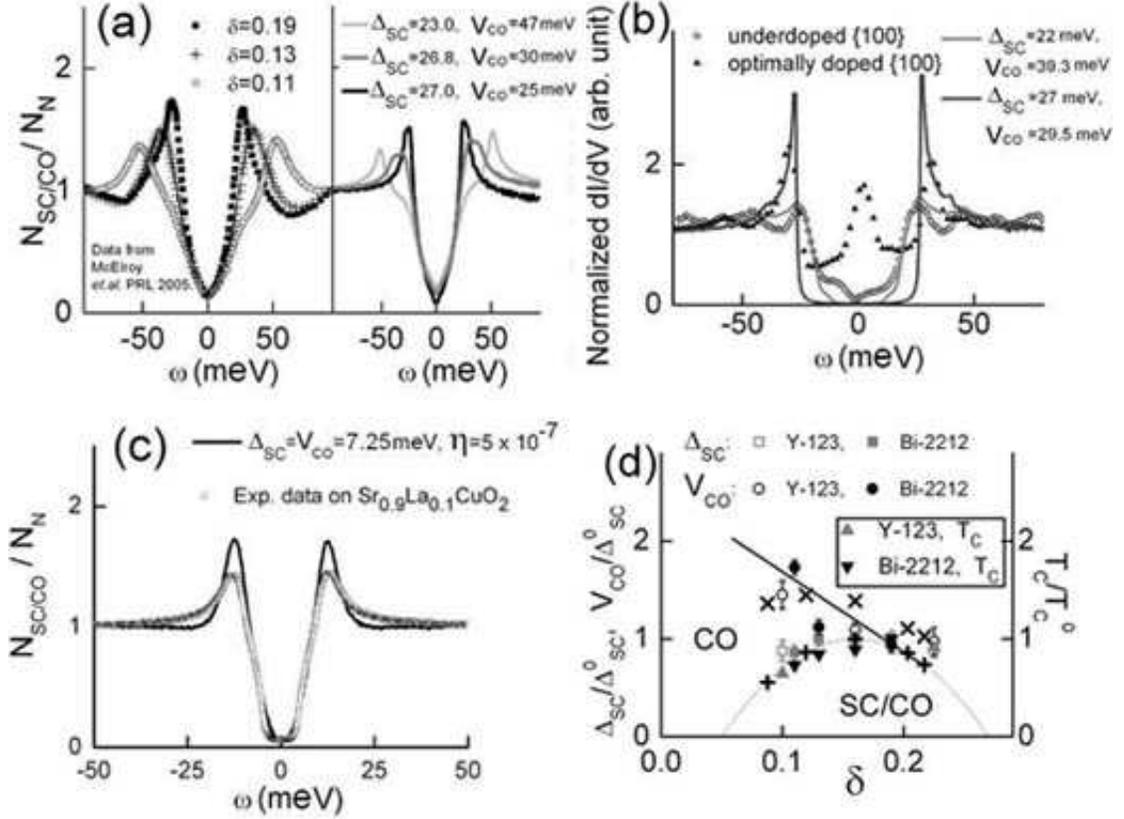}}
\figcaption{(a) Comparison of the c-axis 
($\{ 001 \}$) quasiparticle tunneling spectra on nominally underdoped 
and overdoped Bi-2212 cuprates (left panel, from Ref.~\cite{McElroy05}) with 
theoretical calculations (right panel) using the parameters $\Delta _{\rm SC}$ and 
$V_{\rm CO}$ indicated. (b) Comparison of the anti-nodal ($\{100 \}$) quasiparticle 
tunneling spectra on underdoped ($T_c \approx 60$ K) and optimally doped ($T_c \approx 93$ K) 
Y-123 cuprates (symbols, from Ref.~\cite{Yeh01,Wei98}) with theoretical calculations 
(solid lines). We note that the optimally doped Y-123 was not perfectly aligned 
along the $\{100 \}$ so that some contributions from the zero-bias-conductance-peak 
in the nodal direction~\cite{Wei98} were present in the spectra. 
(c) Comparison of the tunneling spectrum (open squares) on optimally doped La-112 
cuprate~\cite{ChenCT02} with theoretical calculations (solid line). (d) Hole-doping
dependence of $\Delta _{\rm SC}$, $V_{\rm CO}$ and $T_c$ for Y-123 and Bi-2212, where
$\Delta _{\rm SC}$ and $V_{\rm CO}$ are normalized to their corresponding SC gaps at 
the optimal doping, $\Delta _{\rm SC} ^0$, and $T_c$ is also normalized to the value
at the optimal doping $T_c^0$. In addition, the onset temperatures for diamagnetism 
$T_{\rm onset}$ in Bi-2212 with various doping levels from Ref.~\cite{WangY05} are
shown in crosses for comparison, and we note the good agreement between the doping 
dependence of $(T_{\rm onset}/T_c ^0$) and that of ($V_{\rm CO}/\Delta _{\rm SC} ^0$).}
\end{figure}

In addition to accounting for the primary features of doping dependent quasiparticle
tunneling spectra, the notion of coexisting CO and SC can also explain the spatially 
varying local density of states (LDOS) and the corresponding Fourier transformation 
of the LDOS (FT-LDOS) in Bi-2212~\cite{Hoffman02,Vershinin04,McElroy05}. 
As schematically illustrated in Fig.~2(a), the spatially varying LDOS can be 
simulated by designating spatially varying parameters 
$(\Delta _{\rm SC}, V_{\rm CO}, \eta)$ so that the corresponding LDOS 
${\cal N} (\textbf{r}, \omega)$ for a given bulk doping level can reproduce 
the empirical finding. Specifically, we note that overall $\Delta _{\rm SC}$ does not
vary much within each sample, as manifested in Fig. 1(d), while the magnitude of
$V_{\rm CO}$ changes more significantly with the doping level and in space, 
particularly for the underdoped cuprates. We may approximately divide the spectral 
characteristics of Bi-2212 into two types of regions: the ``A''-type region with 
larger $V_{\rm CO}$ accompanied by stronger fluctuations (large $\eta$), and the 
``B''-type region with smaller $V_{\rm CO}$ accompanied by small $\eta$. On the
other hand, $\Delta _{\rm SC}$ remains relatively homogeneous throughout A and B 
regions for a sample of a nominal doping level. The total area of the B-type region 
appears to increases with increasing doping level at the cost of the A-type region. 
Therefore, the appearance of spectral inhomogeneity may be 
largely attribute to variations in $V_{\rm CO}$ and $\eta$, as exemplified in Fig.~2(b) 
for a special case showing spectral variations due to increasing $\eta$ while 
keeping $\Delta _{\rm SC}$ and $V_{\rm CO}$ fixed. We further note that the variations 
in $V_{\rm CO}$ and $\eta$ can effectively result in scattering of quasiparticles. 
Consequently, we expect the FT-LDOS of such quasiparticle spectra in the first 
Brillouin zone to reveal not only excess contributions associated with the CO at the 
wave-vector $\textbf{Q}$~\cite{ChenCT03} but also quasiparticle interference 
effects~\cite{ChenCT03} as the result of spatial inhomogeneity in $V_{\rm CO}$ and 
$\eta$. We further emphasize that the spectral inhomogeneity is NOT the result 
of large variations in $\Delta _{\rm SC}$. Moreover, $V_{\rm CO}$ in Bi-2212 
system apparently exhibit much stronger spatial variations than that in other
cuprates such as Y-123 and La-112, probably due to the stronger two dimensionality 
in Bi-2212.

\begin{figure}[!t]
\centerline{\includegraphics{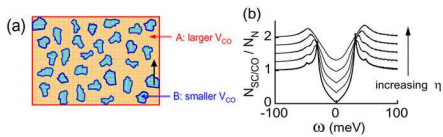}}
\figcaption{(a) Schematic illustration of the LDOS in a slightly underdoped 
Bi-2212 system with coexisting SC and CO. Our theoretical analysis
of experimental data suggests that $\Delta _{\rm SC}$ generally varies slowly
within a given sample, while the apparent spatial variations in the LDOS 
is primarily due to stronger variations in $V_{\rm CO}$ and $\eta$. 
Here we denote the A-type region as areas of larger $V_{\rm CO}$ 
and $\eta$, whereas the B-type region as areas of smaller and nearly homogeneous
$V_{\rm CO}$ and $\eta$. (b) An example of varying spectral characteristics due to
varying quantum phase fluctuation (parameterized by $\eta$) alone while keeping
$\Delta _{\rm SC}$ and $V_{\rm CO}$ constant~\cite{ChenCT06}.}
\end{figure}

\section{Relevant Competing Orders to Cuprate Superconductivity}

While we have demonstrated that CDW and SDW are relevant CO in
accounting for the unconventional low-energy excitations and the lower-energy
pseudogap phenomena, the importance of considering the realistic 
bandstructures and Fermi level in determining whether a CO is relevant to
cuprate superconductivity cannot be over-emphasized. For instance, 
in our consideration of either CDW or SDW as the coexisting CO with SC, 
although the wave-vector $\textbf{Q}$ of the density waves need not be 
commensurate with $(\pi/a)$, maximum effect of CO on the low-energy 
excitations only occurs when $|\textbf{k}|$ and $|\textbf{k}+\textbf{Q}|$ 
are comparable to the Fermi momentum $k_F$ for $\textbf{Q}$ along either
(0,$\pi$) or ($\pi$,0) direction. Specifically, maximum effect of CO
occurs when $|\textbf{Q}| \sim 2k_{Fx}$ for $\textbf{Q}$ along (0,$\pi$) and
$|\textbf{Q}| \sim 2k_{Fy}$ for $\textbf{Q}$ along ($\pi$,0). If we 
relax the condition for $\textbf{Q}$ such that $|\textbf{k}+\textbf{Q}|$ deviates 
substantially from $k_F$, the effect of CO becomes much weakened.
As exemplified in Fig.~3, we compare the effective order parameter
$\Delta _{\rm eff}$ in the first quadrant of the Brillouin zone of $s$-wave SC with 
coexisting CDW (first row) and $d_{x^2-y^2}$-wave SC with coexisting SDW (second row).
For the CO wave-vector $\textbf{Q}$ varying from $|\textbf{Q}| < 2k_F$ (left
panels), $|\textbf{Q}| = 2k_F$ (middle panels), to $|\textbf{Q}| > 2k_F$ (right
panels), we find the strongest CO-induced dichotomy ({\it i.e.}, anisotropic momentum
dependence of $\Delta _{\rm eff}$) for $|\textbf{Q}| = 2k_F$, implying the
maximum effect of the CO on the ground state and the low-energy excitations
of the cuprates if the CO wave-vector is correlated with the Fermi momentum.

\begin{figure}[!t]
\centerline{\includegraphics{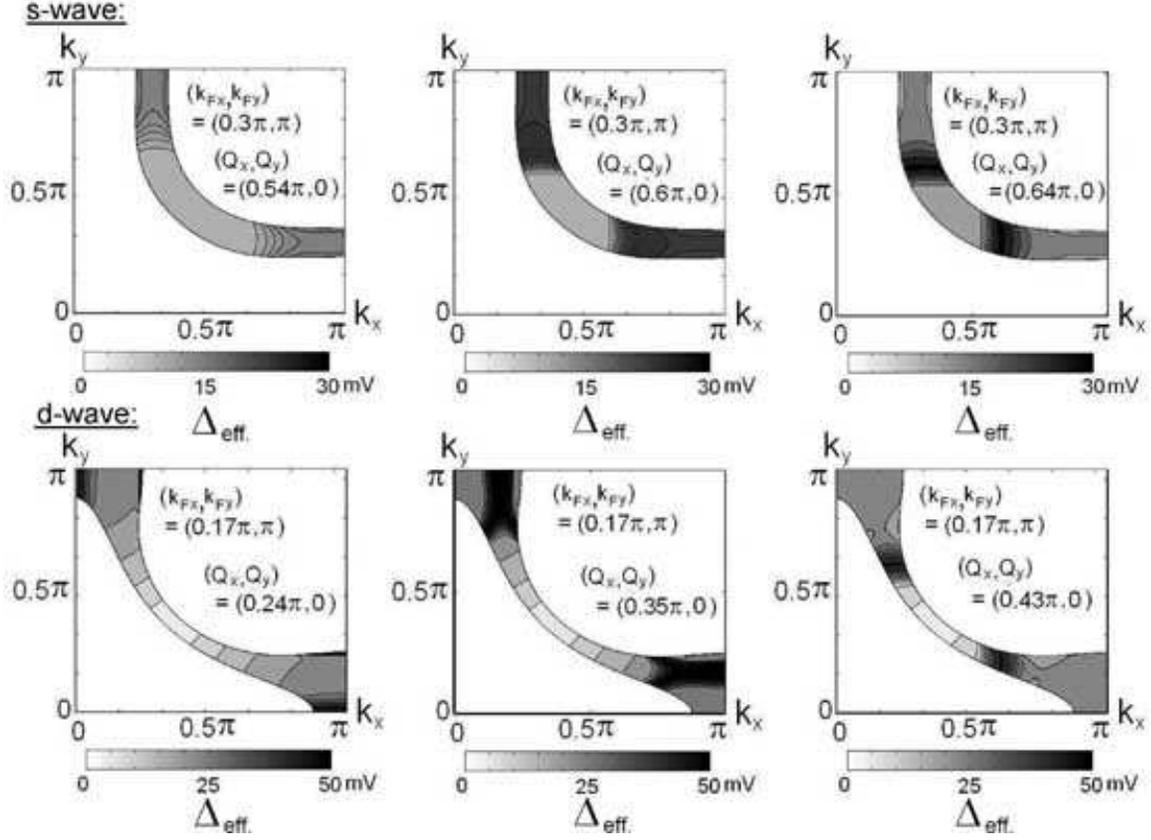}}
\figcaption{Competing order-induced dichotomy in the momentum-dependent
effective order parameter $\Delta _{\rm eff} (\textbf{k})$ of cuprate 
superconductors, where the first row corresponds to $s$-wave SC coexisting with CDW, 
and the second row corresponds to $d_{x^2-y^2}$-wave SC coexisting with SDW. The 
wave-vector $\textbf{Q}$ of the CO along either $(\pi ,0)$ or $(0, \pi )$ 
direction varies from $|\textbf{Q}| < 2k_F$ in the left
panels to $|\textbf{Q}| = 2k_F$ in the middle panels and to $|\textbf{Q}| > 2k_F$ 
in the right panels. Clearly maximum effects of CO occur if $|\textbf{Q}| = 2k_F$,
regardless of the pairing symmetry.}
\end{figure}

In this context, we examine the effect of DDW on cuprate superconductivity and
find that the DDW order parameter does not couple well with the doped cuprates. 
Specifically, its effect is only significant for the nearly nested Fermi surface
(see Fig.~4(a)), which corresponds to a nearly half-filling (and thus 
insulating) condition with bandstructures not representative of the cuprates. 
On the other hand, the contribution of DDW to quasiparticle tunneling spectra 
becomes essentially uncorrelated with experimental observation if we consider
the Fermi surface of a realistic cuprate with a doping level deviating from 
half-filling, as shown in Fig.~4(b). 

\begin{figure}[!t]
\centerline{\includegraphics{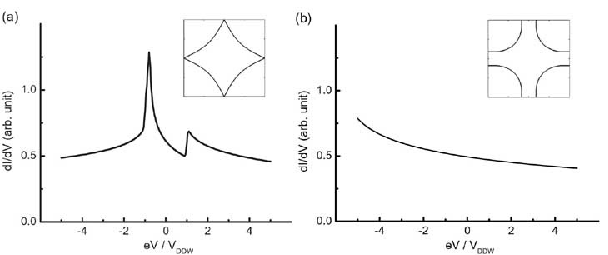}}
\figcaption{Illustration of the effect of DDW on the low-energy excitations
of hole-type cuprates: Quasiparticle DOS due to DDW under (a) nearly 
nested condition~\cite{Bena04} and (b) nearly optimally doped condition 
with realistic cuprate bandstructures~\cite{Norman95}, as 
manifested in the insets for the Fermi surface in the first Brillouin zone. 
Clearly the quasiparticle DOS associated with DDW does not agree with the 
empirical tunneling spectra of cuprate superconductors.}
\end{figure}

\section{Two pseudogap energy scales and the generic phase diagram}

Thus far with the notion of coexisting CO and SC and quantum fluctuations, 
we have successfully accounted for the experimental findings of apparent 
pseudogap phenomena in ARPES and STM of hole-type cuprate superconductors by the 
condition $V_{\rm CO} > \Delta _{\rm SC}$ and for the absence of such phenomena 
in electron-type by the condition $V_{\rm CO} \le \Delta _{\rm SC}$. 
This pseudogap at a lower energy scale ({\it i.e.}, on the order of $\Delta _{\rm SC}$) 
is well correlated with the anomalous Nernst effect~\cite{WangY06} and 
has only been found in hole-type cuprates. In contrast, the higher-energy 
pseudogap (on the order of the Neel temperature) observed in 
optical~\cite{Onose01,Gallais05} and neutron scattering~\cite{Yamada98,Motoyama06} 
experiments on both electron- and hole-type cuprates may have its physical origin 
primarily in magnetism. In the case of Bi-2212 system, the higher-energy pseudogap 
seems correlated with the so-called ``dip-hump'' features in the quasiparticle 
tunneling spectra, whose characteristic energies appear to be related to the
energies of spin fluctuations found in Raman spectroscopy and neutron 
scattering~\cite{Chubukov00}. If the higher-energy pseudogap is indeed 
of magnetic origin and is dynamic in nature, one would expect that the presence 
of an external field can stabilize such a dynamic phase, leading to interesting 
experimental consequences. 

A piece of suggestive evidence associated with this conjecture has been 
recently observed in our tunneling spectroscopic studies of the electron-type 
La-112 system. As shown in the upper panel of Fig. 5(a), for sufficiently small tunneling currents 
the quasiparticle spectrum of optimally doped La-112 exhibits no pseudogap 
above $T_c$~\cite{ChenCT02}, which is as expected because our fitting to the 
data reveals $V_{\rm CO} < \Delta _{\rm SC}$~\cite{ChenCT06}. Interestingly, however, 
if we increase the tunneling currents beyond a critical value ($I_{cr} \sim 45$ nA)
that corresponds to a local magnetic field on the order of a few tens of Tesla, 
we find that the original sharp peaks associated with 
$\omega = \pm \Delta _{\rm eff} \sim \pm \Delta _{\rm SC}$ 
become suppressed, and a second set of peaks at higher energies 
$\omega = \pm V_{\rm PG}$ emerge~\cite{Yeh05a,Yeh05b}, as shown in the lower 
panel of Figure 5(a). The magnitude of $V_{\rm PG}$ is significantly larger
than $\Delta _{\rm SC}$, and is therefore of a different physical origin
from the competing order energy scale $V_{\rm CO}$ and appears to have
been stabilized by the large tunneling currents.

Based on our theoretical calculations and existing experimental observation
of ours and others, we suggest that the generic temperature ($T$) vs. doping 
level ($\delta$) phase diagram of the cuprates is determined by the interplay
of three primary energy scales: $V_{\rm PG}$, $V_{\rm CO}$ and $\Delta _{\rm SC}$,
which correspond to temperature scales of $T_{\rm PG} (\delta)$, $T^{\ast} (\delta)$ 
and $T _c (\delta)$. In the case of hole-type cuprates, generally $V_{\rm CO} > \Delta _{\rm SC}$ for 
a wide range of doping levels, probably due to enhanced charge transfer along the 
Cu-O bonding as the result of significant coupling of the electronic configuration 
to the longitudinal optical (LO) phonons~\cite{Tachiki03}, so that the CO phase occurs 
at $T^{\ast} (\delta)> T_c (\delta)$. In contrast, there is no discernible 
charge transfer due to the LO phonons in electron-type cuprates, and the corresponding 
competing order energy scale becomes much weaker so that $V_{\rm CO} < \Delta _{\rm SC}$. As 
a result, there is no apparent pseudogap associated with electron-type cuprates in the
absence of external magnetic fields~\cite{ChenCT02,Kleefisch01}. On the other hand, 
the higher-energy pseudogap exists in both electron- and hole-type cuprates, 
which may be related to spin fluctuations and thus $V_{\rm PG} \gg \Delta _{\rm SC}$. 
In the language of the slave-boson theory~\cite{LeePA06}, $V_{\rm PG}$ may be thought of
as the spinon pseudogap in the underdoped limit. Hence, our proposed phase diagram in
Fig. 5(b) has effectively unified the seemingly puzzling asymmetric phenomena 
between hole- and electron-type cuprates. In particular, we may consider the 
spinon pseudogap phase determined by the energy scale $V _{\rm PG}$ as the
highly degenerate ``parent phase'' of all cuprates, with antiferromagnetism (AFM), 
SC and CO being the broken-symmetry phases derived from the parent phase.

\begin{figure}[!t]
\centerline{\includegraphics{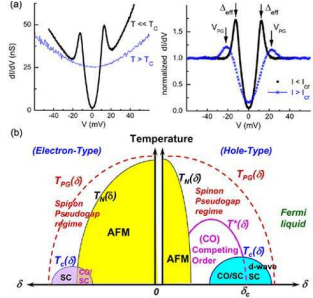}} 
\figcaption{(a) Evidence for two pseudogap energy scales in the electron-type
La-112 system: The absence of the lower-energy pseudogap above $T_c$~\cite{ChenCT02}
shown in the upper panel can be attributed to the condition $V_{\rm CO} < \Delta _{\rm SC}$,
whereas the occurrence of a higher-energy pseudogap $V_{\rm PG}$ induced by large
tunneling currents at $T \ll T_c$~\cite{Yeh05a,Yeh05b} may be attributed a form
of bosonic excitations stabilized by an effective large magnetic fields.
(b) A proposed temperature ($T$) vs. doping level ($\delta$) generic phase diagram 
electron- and hole-type cuprates.}
\end{figure}

\section{Summary}

In summary, we have investigated the physical origin of unconventional low-energy
excitations in cuprate superconductors by considering the effect of coexisting 
competing orders (CO) and superconductivity (SC) and that of quantum fluctuations 
and other bosonic modes on the low-energy charge excitation spectra. 
By incorporating both SC and CO in the bare Green's function and quantum 
phase fluctuations in the proper self-energy, we can consistently account for 
various empirical findings in both hole- and electron-type 
cuprates. Moreover, based on our tunneling spectroscopic studies and
experiments of others, we suggest that there are two energy scales associated 
with the pseudogap phenomena, with the high-energy pseudogap $V_{\rm PG}$ probably of 
magnetic origin and the low-energy pseudogap $V_{\rm CO}$ associated with the 
competing orders. A generic phase diagram of the cuprates that unifies various
asymmetric phenomena in electron- and hole-type cuprates is proposed, in which
the interplay of three primary energy scales $V_{\rm PG}$, $V_{\rm CO}$ and 
$\Delta _{\rm SC}$ determines the ground state phases and the unconventional
low-energy excitations of the cuprates.

\section*{Acknowledgment}
 
The work at Caltech is supported by the National Science Foundation
through Grants \#DMR-0405088, and at the Pohang University
by the Ministry of Science and Technology of Korea. We thank Professor
Xiao-Gang Wen for stimulating discussion.


\end{document}